\documentclass[12pt]{article}

\usepackage{amssymb,epsfig,amsmath}

\graphicspath{{figs/}}

\newcommand{\naiad}{NAIAD}
\newcommand{\ul}{\underline}
\newcommand{\chisq}{$\chi$$^{2}$}

\begin{document}

\begin{center}
\baselineskip=24pt

{\Large \bf Reduction of Coincident Photomultiplier Noise Relevant to Astroparticle Physics Experiments}

\vspace{1cm}

\renewcommand\thefootnote{\fnsymbol{footnote}}
\large{
M.~Robinson\footnote{Corresponding author:  Matthew Robinson E-mail:~matthew.robinson@sheffield.ac.uk},
P.~K.~Lightfoot,
M.~J.~Carson,
V.~A.~Kudryavtsev,
N.~J.~C.~Spooner
}

\vspace{1cm}

Department of Physics and Astronomy, University of Sheffield, Hicks
Building, Hounsfield Road, Sheffield S3 7RH, UK

\vspace{1cm}

\begin{abstract}
In low background and low threshold particle astrophysics experiments
using observation of Cherenkov or scintillation light it is common to
use pairs or arrays of photomultipliers operated in coincidence.  In
such circumstances, for instance in dark
matter and neutrino experiments, unexpected PMT noise events have been observed,
probably arising from generation of light from one PMT being detected
by one or more other PMTs.  We describe here experimental
investigation of such coincident noise events and development of new
techniques to remove them using novel pulse shape discrimination
procedures.  When applied to data from a low background NaI detector
with facing PMTs the new procedures are found to improve noise 
rejection by a factor of 20 over conventional techniques, with
significantly reduced loss of signal events.
\end{abstract}

\end{center}

\vspace{0.2cm}
\noindent {\it Key words:} Photomultiplier noise,
Dark matter, Scintillators

\noindent {\it PACS:} 85.60.H 95.35 29.40.Mc
%PACS - photomultipliers, dark matter, scintillation detectors

\vspace{0.2cm}
\noindent Corresponding author: Matthew Robinson,
Department of Physics and Astronomy, 
University of Sheffield, Hicks Building, Hounsfield Road, 
Sheffield S3 7RH, 
Tel: +44~(0)114 2223553, 
Fax: +44~(0)114 2728079, 
E-mail:~matthew.robinson@sheffield.ac.uk

\pagebreak

\section{Introduction}

Photomultiplier noise can significantly limit the sensitivity of low
background and low threshold astroparticle physics experiments
searching for rare events such as neutrinos or WIMP dark matter using
Cherenkov or scintillator technology.  The drive for low threshold,
which determines, for instance, the minimum energy of neutrino that
can be detected in an astrophysical neutrino detector, requires
detection of the lowest possible number of photons in the target
medium.  This translates to design geometries that use many
photomultipliers to view the medium, operated in coincidence.  Such
designs provide the necessary good light collection but furthermore,
because of the coincidence operation, provide an effective means of
suppressing certain types of well known PMT ({\bf P}hoto{\bf
m}ultiplier {\bf T}ubes) noise, in particular
single-photoelectron (SPE) noise and after-pulses.  For these types of noise
there is no causal relation between noise generated in a given tube
and that in any other tube.

 However, operation of these designs in which PMTs are effectively
 light coupled together, makes them vulnerable to other forms of PMT
 noise, in particular to effects often ascribed to ``dynode
 glow'' \cite{bib:noise1,bib:noise2}.  This is believed to be due to
 light emitted from the dynode stack under electron bombardment
 being scattered and reflected back to the photocathode.  This can
 produce noise in the host tube but additionally the photons produced
 can in principle also be observed by all light-coupled tubes.
 Such a coincident event is not suppressed by the
 coincidence operation and may mimic a signal event.  Significant
 phenomena of this sort have been observed in the SNO detector, where
 they have been termed ``flashers''.  These are seen at $\sim$1000/day above
 an 18 hit threshold comprising relatively prompt flashes of light
 attributed to discharges from the dynodes \cite{bib:wark}.  In dark matter
 detectors based on NaI, liquid Xe and other scintillators where
 high sensitivity and low background is needed, a likely related class
 of events sometimes termed ``step'' events is seen \cite{bib:quenby,bib:robinson}. 
 The drive for lower thresholds and lower backgrounds means that such
 unusual and rare events will become an increasingly significant
 source of systematic error in these and other planned PMT-based low
 background detectors.

In this paper we describe investigation of such coincident PMT noise
events and the development of new techniques to identify and suppress
these events.  The work was undertaken using apparatus at the Boulby
Underground Laboratory comprising a low background shielding castle
within which can be mounted pairs of face-to-face low background
photomultipliers with and without
scintillator in various configurations (see sec 2).  The apparatus used is based on that of the
modules of the \naiad\ (Sodium Iodide ({\bf NaI}) {\bf A}dvanced {\bf
D}etector) experiment \cite{bib:naiad}.  The \naiad\ experiment is a direct search for
WIMP (\underline{W}eakly \ul{I}nteracting \ul{M}assive \ul{P}article)
dark matter, comprising an array of 8 shielded low background NaI(Tl)
crystals viewed by pairs of photomultipliers \cite{bib:vitaly}.  WIMP search experiments
use various techniques to distinguish between nuclear recoil events,
expected from WIMP interactions, and background electron recoil
events.  In the case of NaI(Tl) crystals, such as used in \naiad, and
several other experiments, this discrimination is achieved
using pulse shape analysis to identify nuclear
recoils \cite{bib:naiad,bib:vitaly,bib:zeplin,bib:psdother1,bib:psdother2,bib:psdother3}.
This is possible because in NaI(Tl), for instance, the
scintillator time constant for nuclear recoil events is shorter
(typically 170-200 ns) than for electron recoils (typically 240-300 ns).

Dark matter limits are set using this technique by determining
statistically the maximum number of recoil events that may be present
in a dataset.  However, as with many such low background experiments
it is crucial to have sensitivity at the lowest possible energy, to
probe the maximum range of initial particle energy or mass. 
For this reason the coincidence threshold is usually set
low, for instance at approximately 1 keV in \naiad, allowing a sensitivity to WIMP
masses in principle below 100 GeV. In such experiments the rate of PMT
noise events is greatly reduced by using coincidence counting which
demands that each PMT should see a minimum signal for an event to be
accepted.  However, any coincident PMT noise remains and strategies to
reduce this (rather than just the SPE or non-causal noise) then
become of major importance.

\begin{figure}[htb]
\begin{center}
\epsfig{figure=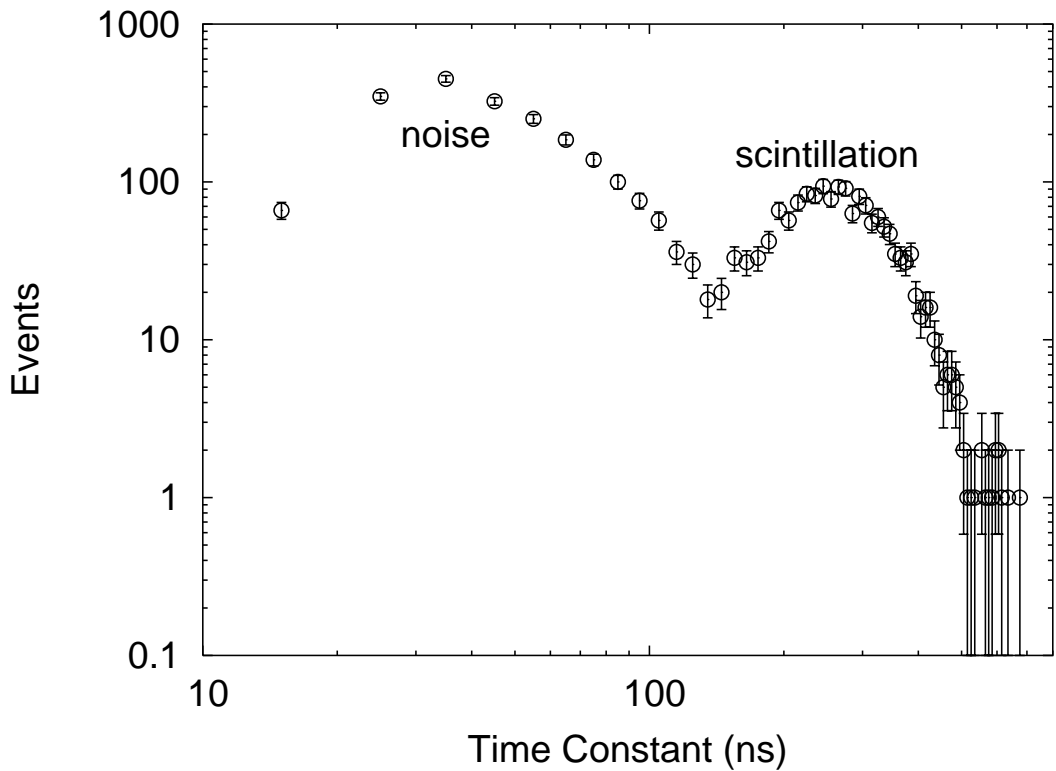,width=15cm}
\caption{Typical \naiad\ time constant distribution for the energy
range 4-6 keV showing
the encroachment of the noise into the scintillation region in data
selected by asymmetry cuts}
\label{fig:tcd}
\end{center}
\end{figure}

In low background experiments using optically coupled PMTs such as this,
some suppression of coincident noise can be achieved by the following two
techniques: (a) additional pulse shape analysis of the events, and (b)
asymmetry cuts.  The former technique is based on the observation that
the distribution of time constants of low energy PMT coincident
noise events can be fitted to an exponential function.  This has a
faster characteristic mean decay time than that of most scintillator
events (though slower than the normal PMT response time).  This
distribution can be fitted and then subtracted from the data,
even though the distribution overlaps with the distribution of time
constants for scintillation events.  Figure \ref{fig:tcd} shows a
typical low energy time constant distribution for NaI including this
noise feature.  The asymmetry cut technique is based mainly on
comparing the characteristics of coincident events, such as amplitude,
absolute start time of the pulse and pulse shape \cite{bib:naiad,bib:vitaly,bib:ahmed}.  The
assumption used is that coincident noise seen in the tubes is 
due to light emitted from one tube being seen by one or more other tubes, 
as suggested occurs in SNO and dark matter
experiments \cite{bib:wark,bib:robinson}.  In this case, unlike the situation for genuine signal
events, the characteristics are likely to be asymmetric, for instance
the pulse in one PMT being of much larger amplitude than the other, or with
faster rise time.

Although these techniques are useful they both have clear
drawbacks.  The coincident noise subtraction technique, although
powerful,  assumes that the exponential fit of short time constant events may be extrapolated to higher time
constants, through the data events. This assumption introdues systematic errors into the results.  Such a subtraction also necessarily
entails the possible rejection of genuine events, requiring the introduction
of an efficiency factor. The asymmetry cuts can not remove
coincident noise events which give pulses in the tubes of similar
characteristic. The asymmetry technique also requires operation in
coincident mode, which may not always be possible or desirable.  For
dark matter searches any PMT noise events
misidentified as nuclear recoil events reduce the effectiveness of
pulse shape discrimination analysis.

Based on the drawbacks above there is a need to develop new
techniques to reject coincident noise events with fit time constants
comparable to signal events.  We describe here new work in this area. 
Firstly, the apparatus used to acquire and characterise samples of low
background coincident PMT noise data is outlined.  This is followed by
details of how the data were analysed and compared with data taken
with similar configurations but with scintillator present.  We then
discuss the development of analysis techniques designed specifically to
identify and reject coincident PMT noise.  Finally, we will show
comparisons between the effectiveness of the new procedures and more
traditional techniques.

\section{Experimental setup and data}

\begin{figure}[htb]
\begin{center}
\epsfig{figure=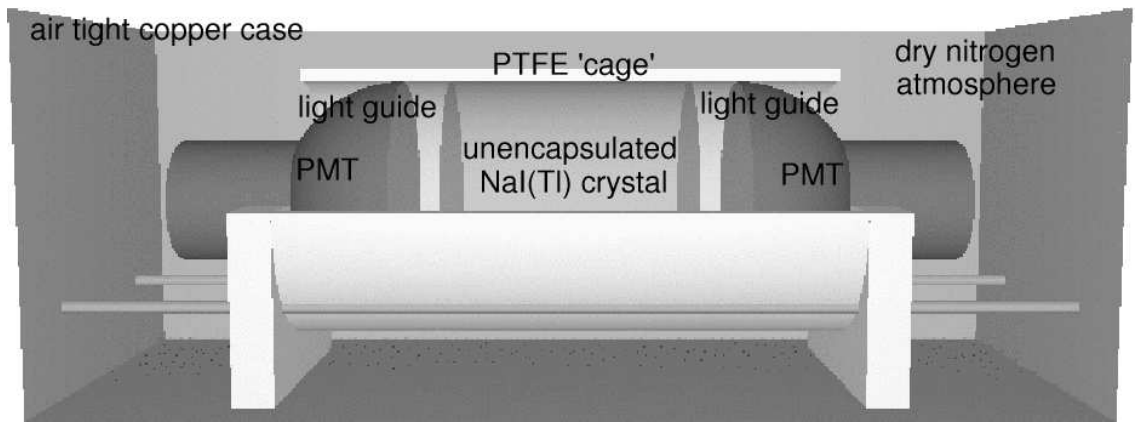,width=15cm}
\caption{Diagram of standard \naiad\ detector module}
\label{fig:naiad}
\end{center}
\end{figure}

All PMT noise measurements were performed at the Boulby
Underground Laboratory at 1070 m depth, using the apparatus shown in
Figure \ref{fig:naiad}.  The set up comprises a shielding castle,
incorporating 15 cm of low background lead, inside of which is a
further 10 cm layer of copper shielding.  The castle was supplied
with an automatic source dropper to allow remote calibration of the
scintillator using $\gamma$-ray sources ($^{57}$Co and $^{60}$Co).  A copper
support structure allows 5 inch PMTs to be mounted face to face in a
variety of configurations within the castle.  EMI 9390 5 inch PMTs were
used throughout.  Four arrangements were used in the present work:

(i) PMTs optically coupled to each other through two 14 cm diameter
cylindrical quartz light guides each 5 cm thick.

(ii) PMTs optically coupled directly with no intervening light guide.

(iii) PMTs coupled together but with the quartz lightguides separated
and coupled through a 5 cm long NaI(Tl) crystal of 14 cm diameter.

(iv) PMTs coupled directly face to face but with an opaque barrier to
stop direct light transmission between the tubes.

Data were collected using an in-house LabView based data aquisition
system \cite{bib:robinson},  the signals from each PMT being first passed
through an integrating buffer.  The integrated signals were fed to
discriminator units set to trigger when signals exceed a threshold of
10 mV, equivalent to 4 photoelectrons.  The logic signals from the
discriminator units were fed to a coincidence unit which triggers a
digital oscilloscope when signals arrive from both discriminators. 
When the oscilloscope was triggered, the integrated signals from both
PMTs were digitised and stored, each digitisation point representing
10 ns elapsed time.  These signals were then read through a GPIB
(\ul{G}eneral \ul{P}urpose \ul{I}nterface \ul{B}us) system into a
computer which wrote the digitised pulses to a data file for
later analysis.

\begin{figure}[htb]
\begin{center}
\epsfig{figure=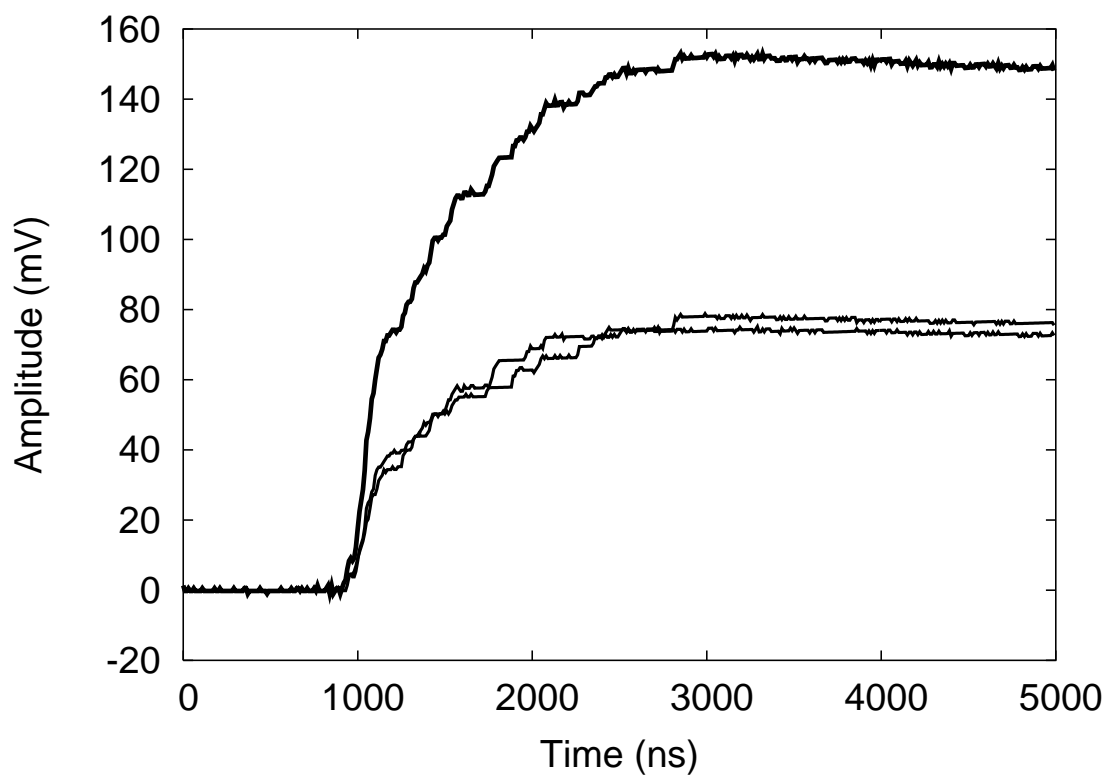,width=15cm}
\caption{Example scintillation event from \naiad.  The 2 lower curves
show the integrated signals from the individual PMTs.  The upper
curve shows the sum of these signals.}
\label{fig:realevt}
\end{center}
\end{figure}

Figure \ref{fig:realevt} shows an example gamma pulse from the
system when configured with set-up (iii) with the light guides and
NaI(Tl) crystal in place.  Two integrated pulses, one from each tube,
are seen, plus the summed pulse.  In such integrated pulses single
photoelectrons are observed as small steps in the pulse.  A conventional
pulse shape analysis, such as used in dark matter searches, is
performed by fitting the summed integrated signal to an exponential
function:

\begin{equation}
\label{eq:exp}
f(A, t,t_0,\tau)=A(1-e^\frac{(t_0-t)}{\tau}),
\end{equation}
\vspace{5mm}

where $A$ is the amplitude of the integrated pulse, which is
proportional to the number of photoelectrons, $t_0$ is the start time
of the rise of the integrated pulse, and $\tau$ is the exponential fit
time constant.  Further details of this analysis are given in
\cite{bib:naiad,bib:vitaly}.  

\begin{figure}[htb]
\begin{center}
\epsfig{figure=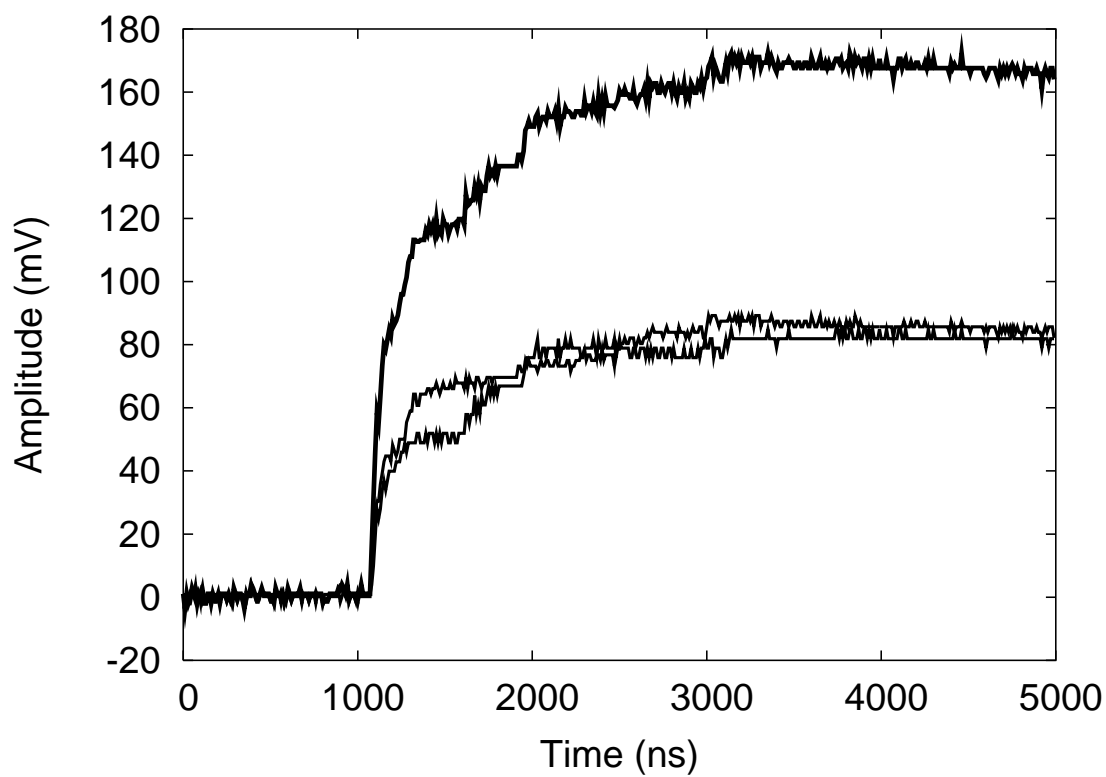,width=15cm}
\caption{Event collected in noise measurement experiment mimicking
scintillation.  As in Figure \ref{fig:realevt}, the 2 lower curves
show the integrated signals from the individual PMTs.  The upper
curve shows the sum of these signals.}
\label{fig:noiseevt}
\end{center}
\end{figure}

A pulse such as shown in Figure \ref{fig:realevt} is
most likely due to scintillation.  Tests with the apparatus configured in an
identical manner but without the crystal in place (set-up (i)) reveal,
as shown by the example pulse in Figure \ref{fig:noiseevt}, that
coincident PMT noise pulses may lead to pulse shapes with similar characteristics.   Tests without the
light guides in place (set-up (ii)) demonstrated that similar events
are still present and so confirmed that the quartz itself is not
responsible for the events.
Events of this kind would not be cut by conventional
asymmetry cuts as the amplitudes and shapes in both tubes are similar.
Identical runs but with an opaque card in place to block
direct passage of light between the tubes (set-up (iv)) confirmed that
in this case no coincident events are observed.   All these results together confirm the
suspicion that slow ($>$100~ns) coincident noise events exist and that they are
due to light generated in one tube being recorded by both tubes.

\begin{figure}[htb]
\begin{center}
\epsfig{figure=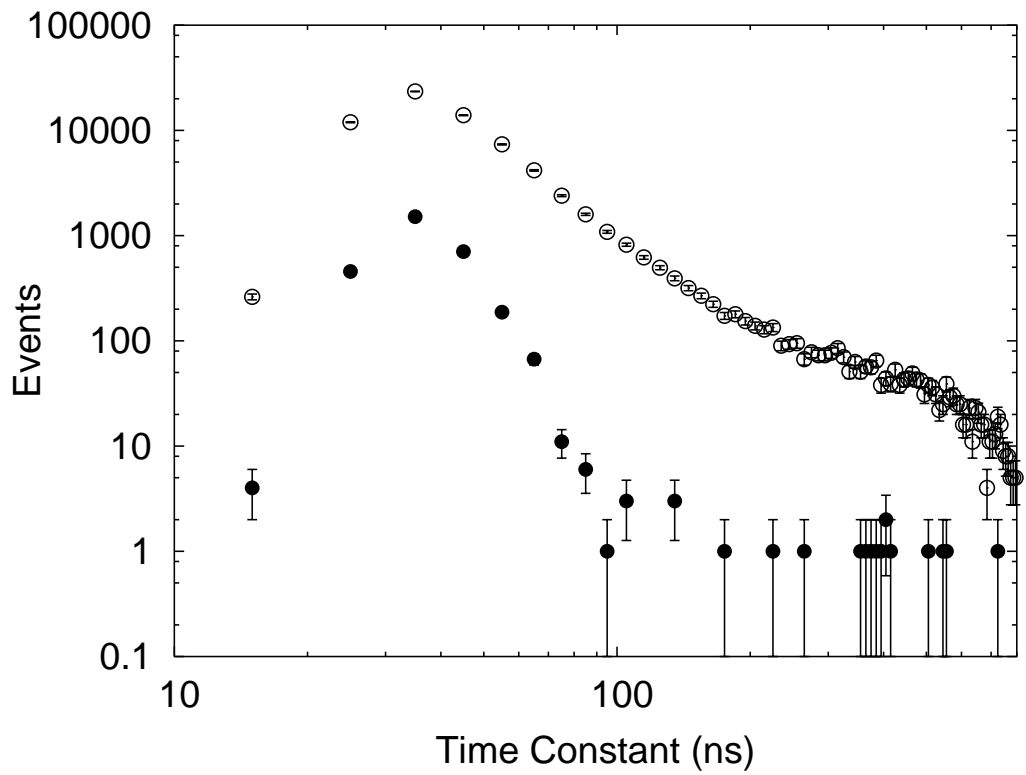,width=15cm}
\caption{Time constant distribution from noise experiment data
corresponding to the energy range 4-6 keV; open
circles represent uncut data, filled circles represent quality cut data}
\label{fig:noisetcd}
\end{center}
\end{figure}

In accordance with NaI analysis procedures developed by the UKDM
Collaboration \cite{bib:naiad,bib:vitaly,bib:ahmed}, data recorded from the experiments
here were binned in 2-d histograms of pulse height (energy) and fit
time constant.  Figure \ref{fig:noisetcd} (upper points) shows a
typical extracted 1-d time constant distribution for the coincident
noise events taken without scintillator for a pulse amplitude band
equivalent to 4-6 keV (set-up (i)) with no cuts.  With the
scintillator in place calibration with $^{57}$Co (122 keV gamma line)
is used to determine the energy scale.  Comparison with the results
with no scintillator then allows an effective energy scale to be
determined when no scintillator is present.  For the tests with scintillator
a $^{60}$Co source was used to generate low energy compton
electron recoils via compton scattering in the crystal with a broad
spectrum of energies.  Since the rate of scintillation events in such
calibration runs is much higher than that of any noise events the data
may be used as a pure sample of scintillation events.  By comparing
noise events from data collected with no scintillator and
scintillation events from data collected with no noise, with similar
amplitude and time constant, it is possible to identify features which
are much more common in the noise events than in scintillator events
and thereby to identify noise events in real data.  The rate fall off
with increasing time constant seen in Figure \ref{fig:noisetcd} (upper
plot), has the characteristic exponential form that is used as a basis
for the conventinal noise subtraction technique \cite{bib:ahmed}. 
This form is seen clearly in the NaI data of Figure \ref{fig:tcd} at
low time constant.

\section{Investigation and improved reduction of noise}

Typical asymmetry cuts used to reject noise from \naiad\ data, reject
events where the fit time constants of pulses from each PMT differ by
100~ns or more, or where the start times of the pulses differ by
100~ns or more, or where the energy associated with the pulses differs
by 40\% of the combined energy or more \cite{bib:robinson,bib:ahmed}. 
Cuts based on these parameters represent the current best procedures
but, as discussed above, do not adeqately deal with the coincident
noise phenomena.  To improve the situation and develop better
rejection strategies the characteristics of noise pulses were examined
more closely.  This study revealed, in particular, that a large
population of the observed smaller multi-photoelectron noise events
when fitted to exponentials had very short time constants
($<$100~ns).  Based on this, the assumption was made that the noise
events observed with longer time constants are combinations of such ``step'' events each separated by a few tens of ns.  These steps may
have some causal relationship but are not the result of a true
exponential decay as is the case for true scintillation pulses. 
Working from this assumption, various quantities relating to the
quality of fits to exponentials were investigated, to determine which
could be used to distinguish the multi-step noise events from
scintillation events with a similar decay profile.  Cuts
defined based on these quantities have been termed ``quality cuts''.

The most obvious quality cut arises from measuring the \chisq\ per
degree of freedom:

\begin{equation}
\label{eq:chi2}
\chi^2=\sum_{i=n_0}^{n_1}{\frac{(f(A, t_i,t_0,\tau)-D_i)^2}{\sigma_i^2}},
\end{equation}

where $n_0$ is the first data point included, $n_1$ is the last data
point included, $f(t_i,t_0,\tau)$ is the fit as described in Equation
\ref{eq:exp}, $t_i$ is the time at digitisation point $i$, $D_i$ is
the measured amplitude at that digitisation point and $\sigma_i$ is
the uncertainty on that amplitude.
 
Unfortunately, although a least squares method
is used to fit the pulses to an exponential, calculating the \chisq\
is not straightforward because the fit is performed on integrated
pulses and the uncertainties on each digitised pulse point are not
defined.  To deal with this problem, a value for uncertainty is chosen
for each pulse and applied to all digitised points within that pulse.
This uncertainty is chosen to be proportional to the square root of
the total number of photoelectrons the pulse represents and
is normalised to yield an average \chisq\ per degree of freedom of 1.0 for
scintillation events of all energies.

The rise of an integrated pulse represents the arrival of photons at
the PMT, but the flat part of the pulse following the rise does not. 
Since no new information is associated with this part of the digitised
pulse, tests were made with evaluation of the fit quality within
limited time regions of the pulse.  The typical scintillation time
constant is 250-300 ns and it was found that calculation of \chisq\
per degree of freedom using the first 50~ns, 50~to~100~ns
or 100~to~500~ns of pulse data points give quantities which are useful
in identifying noise events.  The \chisq\ calculated using the full pulse
was also found to be useful.  Usually one or more of these particular
quality parameters is typically larger for a given noise event than
for similar scintillation events.  However, no one parameter set
appears sufficient alone to identify all noise events.  Noise may be
identified by some of the parameters but missed by others.

To address this, a further new quality cut technique was devised which
we have called ``steppiness''.  Steppiness ($\Gamma$) provides a means to
quantify the tendency of the digitised PMT coinicident noise pulse to
be made up of multi-photoelectron steps.  It is described by the
following equations:

\begin{equation}
\label{eq:steppiness}
\Gamma=\frac{\sum_{i=1}^{N-n}{\frac{\mu_i}{f'(A, t_i,t_0,\tau)}}}{N-n}
\end{equation}

\vspace{0cm}
\begin{displaymath}
{\rm where\ \ \ }\mu_i = \left \{\begin{tabular}{ll}
$\nu_i-\nu_{i-1}$ : $\nu_i-\nu_{i-1} \ge \nu_{crit}$ \\
$0$ : $\nu_i-\nu_{i-1} < \nu_{crit}$
\end{tabular}
\right.
\end{displaymath}

\begin{displaymath}
{\rm and\ \ \ }\nu_i=\frac{\sum_{j=i}^{i+n-1}{D_j}}{n}.
\end{displaymath}

Calculation of steppiness for a particular pulse proceeds as follows:
each digitised point ($D_j$) is converted to an average over $n$
points starting from $D_i$ and ending at $D_{i+n-1}$.  $N$
digitisation points are used from the start of the pulse ($t_0$) to
the end of the digitisation range.  It was found that $n=3$
(corresponding to a 30~ns time window) gives optimum results.  This
value is labelled $\nu$.  $\mu$ is the change in $\nu$ from one
digitisation point to the next.  If the change in $\nu$ is less than
that for the arrival of 2 photoelectrons ($\nu_{crit}$), calculated
based on a single photoelectron calibration of the PMT, then $\mu$ is
set to zero.  This was done because steps of one photoelectron are to
be expected throughout the pulse and such a step is not an indication
of pulse quality.  The sum $\mu$ at each digitisation point divided by
the gradient of the fit at the same point is calculated and converted
to an average to acquire the value of steppiness.  The gradient is
used as a measure of how reasonable it is to have a step of certain
size at each point and serves to weight the values.

\begin{figure}[htb]
\begin{center}
\epsfig{figure=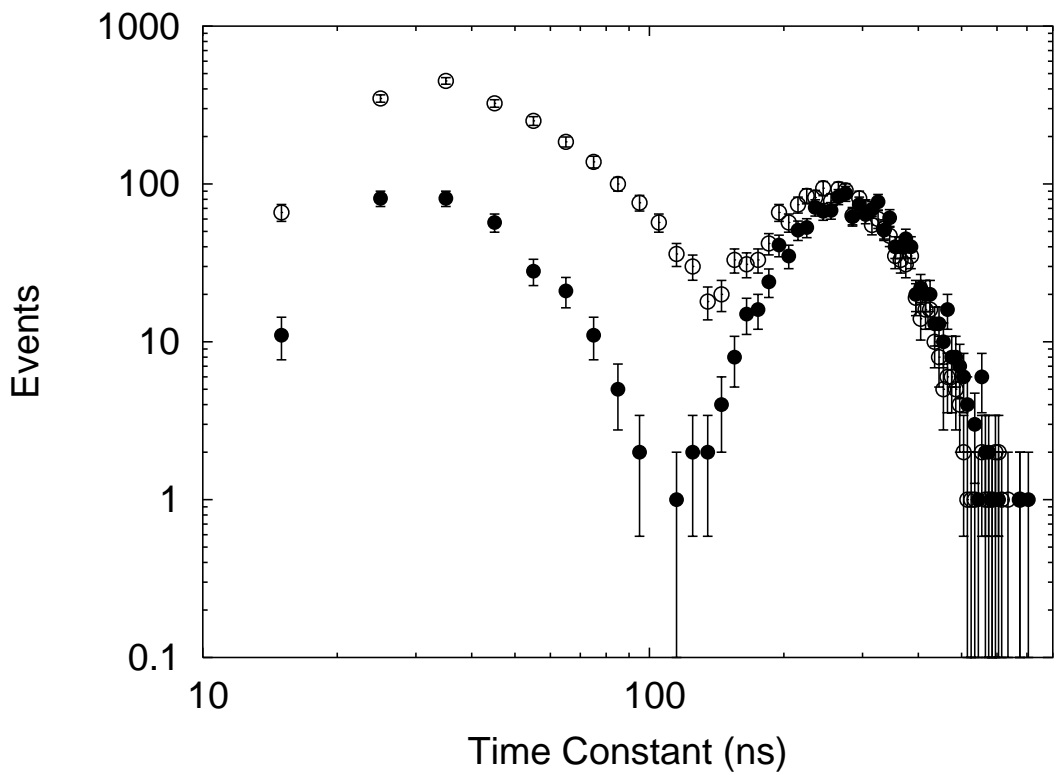,width=15cm}
\caption{Time constant distribution showing relative effectiveness of
quality cuts (filled circles) versus traditional cuts (hollow circles)
in the energy range 4-6 keV}
\label{fig:datatcd}
\end{center}
\end{figure}

The steppiness quantity and the full \chisq\ quantity were found to
best identify noise events which have been fitted with a long time
constant ($>$300 ns) whereas the partial \chisq\ quantities were found
to work best on faster events ($<$300 ns).  Together, it has been
found that these quantities may be used to efficiently identify the
full range of noise events with time constants in the scintillation
region.  This is illustrated in Figure \ref{fig:noisetcd}. 
Here the effect of the quality cuts on a sample of pure
PMT coincident noise data is shown for events taken without
scintillator, and for 
pulse heights equivalent to 4-6 keV. It can be seen that the noise
suppression factor is $>$1000 above 70 ns and still a factor 100
at lower time constants.  Figure \ref{fig:datatcd} shows the effect of
quality cuts on real experimental data (lower points) compared with
the effect of using conventional asymmetry cuts (upper points).  An
excellent reduction in noise can be seen in the critical
scintillation-like region of 30-200 ns.  The improvement at 100 ns
being about a factor of 20.  For NaI dark matter searches the key region of 
interest in a plot such as this is the left side of the 
main peak at 100-200 ns, the region expected for nuclear recoil 
events.  The remarkable noise reduction and clarification of the 
electron signal here can be expected to improve sensitivity and 
control of systematic errors that result from usual noise subtraction 
techniques.

In addition, calculations were made of \chisq\ over what is expected to
be the flat region of the pulse at $(t-t_0)>$$500$ ns (see Figure
\ref{fig:realevt}).  This quantity was found to be very effective for
identifying events containing either PMT after-pulse signals or noise
events within scintillation events.  These events are rejected from
\naiad\ because the presence of the after-pulse may affect the
accuracy of the least squares fitting routine.

Any loss of genuine scintillation or Cherenkov events from experimental
data due to application of noise cuts clearly reduces experiment
sensitivity.  In the case of
scintillators such as NaI using pulse shape analysis the important
parameters are the proportion of electron recoils lost due to cuts
and, for dark matter, the propotion of nuclear recoil events lost. 
Various test runs were performed to examine these efficiency losses
using gamma and neutron sources.  In the current apparatus, due to the
saturation of the DAq at high rates, the $^{60}$Co calibration data
has a much lower PMT noise content than normal low background data.  The
scintillation event rate for such data is much higher (typically
50~s$^{-1}$ within the energy range 4-15~keV compared to less than
1~s$^{-1}$ for data collected with no source) but the PMT noise rate
remains constant.  The fraction of events cut from such data therefore
gives a good indication of the efficiency of the cuts.  Table
\ref{tab:efficiency} shows the efficiency of the newly developed cuts
calculated in this manner compared to the efficiency of the
traditional asymmetry cuts calculated by applying the cuts to
$^{60}$Co calibration data collected in the Boulby mine underground
laboratory at a controlled crystal temperature of 8$^\circ$C. The
quality cuts have been optimised to remove all noise with time
constants in and around the scintillation region.  Comparing,
for instance, the energy band at 6-8 keV (critical in dark matter
searches) the loss of valid events due to asymmetry cuts is no greater than 16\%. The
quality cut process yields a slightly smaller loss, lower by typically 4\%.
   
For NaI dark matter experiments the goal is to search for nuclear
recoil events in experimental data, so efficiency in terms of fraction of
rejected nuclear recoil events is also important.  Neutron source
calibration data collected on the surface has a very low noise content
in much the same way as $^{60}$Co calibration data.  Such data were
used to measure the efficiency of the cuts.  Table
\ref{tab:sefficiency} compares cut efficiencies calculated by applying
the cuts to $^{60}$Co calibration data collected on the surface at
room temperature.  Table \ref{tab:nefficiency} compares cut efficiency
calculated by applying the cuts to neutron calibration data collected
on the surface at room temperature.  These two tables show that the
efficiencies for electron recoil events are similar to those for
nuclear recoil events.  Again the quality cuts cause slightly less 
loss of events, by typically 4-7\%.

\section{Conclusions}

An experiment to acquire a sample of pure PMT coincident noise data
was performed using a low background shielding array underground at
the Boulby facility.  These data were analysed and parameters
developed to distinguish and thereby cut this noise from real data. 
The newly developed cuts have been shown to be more effective at
removing noise from experimental data and more efficient in preserving
scintillation events compared to the asymmetry cuts currently used. 
In the typical scintillator region of around 100 ns the noise
rejection is improved by a factor of 20.  These cuts have further been shown to
preserve nuclear recoil scintillation events with the same efficiency
as electron recoil scintillation events and in both cases with 
slightly better efficiency than possible with assymetry cuts (by 
about 4-7\%).  Similar quality cuts may prove
effective in a range of experiments using PMTs required to work near 
threshold.  Furthermore, since the new technique no longer relies on 
measuring coincidence asymmetry, it can in principle be applied to 
single PMT set-ups.  The cuts developed here will be used in analysing 
data from \naiad\ and are expected to produce a significant
improvement in the sensitivity of the experiment and control of 
systematic error that previously resulted from noise subtraction.

\section{Acknowledgements}

The authors would like to thank the members of the UK Dark Matter
Collaboration for their valuable assistance and advice.  
We are grateful to the Particle Physics and
Astronomy Research Council for financial support and to Cleveland Potash
Limited for their assistance.
M. Robinson would also like to thank Hilger Crystals for their support
of his PhD work.

\pagebreak

\begin{table}
\begin{center}
\caption{Effect of cuts on scintillation events in underground $^{60}$Co calibration data} 
\vspace{3mm}
\begin{tabular}{*{4}{|c}|}
\hline
\label{tab:efficiency}
energy range & without cuts & quality cuts & asymmetry cuts  \cr
(keV) & (events) & (events remaining) & (events remaining)  \cr
\hline
4-6 & 1076 & 911 (85\%) & 834 (78\%) \cr
6-8 & 1081 & 958 (88\%) & 910 (84\%) \cr
8-10 & 1106 & 1069 (97\%) & 1010 (91\%) \cr
10-12 & 1081 & 1062 (98\%) & 1018 (94\%) \cr
\hline
\end{tabular} 
\end{center} 
\end{table}

\begin{table}
\begin{center}
\caption{Effect of cuts on scintillation events in surface $^{60}$Co calibration data} 
\vspace{3mm}
\begin{tabular}{*{5}{|c}|}
\hline
\label{tab:sefficiency}
energy range & without cuts & quality cuts & asymmetry cuts  \cr
(keV) & (events) & (events remaining) & (events remaining)  \cr
\hline
4-6 & 1521 & 1240 (82\%) & 1209 (79\%)\cr
6-8 & 3872 & 3619 (93\%) & 3380 (87\%) \cr
8-10 & 7074 & 7024 (99\%) & 6532 (92\%) \cr
10-12 & 10442 & 10382 (99\%) & 9918 (95\%) \cr
\hline
\end{tabular} 
\end{center} 
\end{table}

\begin{table}
\begin{center}
\caption{Effect of cuts on scintillation events in surface neutron calibration data} 
\vspace{3mm}
\begin{tabular}{*{5}{|c}|}
\hline
\label{tab:nefficiency}
energy range & without cuts & quality cuts & asymmetry cuts  \cr
(keV) & (events) & (events remaining) & (events remaining)  \cr
\hline
4-6 & 6488 & 5458 (84\%) & 5234 (81\%) \cr
6-8 & 13179 & 12477 (95\%) & 11887 (90\%) \cr
8-10 & 20554 & 20413 (99\%) & 19359 (94\%) \cr
10-12 & 27019 & 26876 (99\%) & 26097 (97\%) \cr
\hline
\end{tabular} 
\end{center} 
\end{table}

\pagebreak

\end{document}